\documentclass[conference]{IEEEtran}
\IEEEoverridecommandlockouts
\usepackage{cite}
\usepackage{amsmath,amssymb,amsfonts}
\usepackage{algorithmic}
\usepackage{graphicx}
\usepackage{textcomp}
\usepackage{xcolor}
\usepackage{todonotes}
\usepackage{float}
\usepackage{geometry}
\usepackage{mathrsfs}
\usepackage{cite}
\usepackage{balance}
\usepackage{textcomp}
\usepackage{subcaption}
\ifCLASSINFOpdf
\else
\usepackage[dvips]{graphicx}
\fi
\newtheorem{remark}{Remark}

\usepackage{algorithm}      
\usepackage{algorithmic}

\usepackage{varwidth}

\begin{document}
\newgeometry{left=19.1mm,bottom=19.1mm,top=25.4mm,right=19.1mm}

\title{Impact of Information Flow Topology on Safety of Tightly-coupled Connected and Automated Vehicle Platoons Utilizing Stochastic Control}



\author{Mahdi Razzaghpour$^*$, Sahand Mosharafian$^{\dag}$, Arash Raftari$^*$, Javad Mohammadpour Velni$^{\dag}$, Yaser P. Fallah$^*$\\ \\
$^*$Dept. of Electrical \& Computer Engineering, Univ. of Central Florida, Orlando, FL\\
$^{\dag}$School of Electrical \& Computer Engineering, Univ. of Georgia, Athens, GA 
\thanks{This research was supported by the National Science Foundation under grant numbers CNS-1932037 and CNS-1931981.}}
\maketitle

\begin{abstract}

Cooperative driving, enabled by Vehicle-to-Everything (V2X) communication, is expected to significantly contribute to the transportation system's safety and efficiency. Cooperative Adaptive Cruise Control (CACC), a major cooperative driving application, has been the subject of many studies in recent years. The primary motivation behind using CACC is to reduce traffic congestion and improve traffic flow, traffic throughput, and highway capacity. Since the information flow between cooperative vehicles can significantly affect the dynamics of a platoon, the design and performance of control components are tightly dependent on the communication component performance. In addition, the choice of Information Flow Topology (IFT) can affect certain platoon's properties such as stability and scalability. Although cooperative vehicles’ perception can be expanded to multiple predecessors’ information by using V2X communication, the communication technologies still suffer from scalability issues. Therefore, cooperative vehicles are required to predict each other's behavior to compensate for the effects of non-ideal communication. The notion of Model-Based Communication (MBC) was proposed to enhance cooperative vehicle's perception under non-ideal communication by introducing a new flexible content structure for broadcasting joint vehicle's dynamic/driver's behavior models. By utilizing a non-parametric (Bayesian) modeling scheme, i.e., Gaussian Process Regression (GPR), and the MBC concept, this paper develops a discrete hybrid stochastic model predictive control approach and examines the impact of communication losses and different information flow topologies on the performance and safety of the platoon. The results demonstrate an improvement in response time and safety using more vehicles' information, validating the potential of cooperation to attenuate disturbances and improve traffic flow and safety.

\end{abstract}


\section{Introduction} \label{sec::intro}

\noindent The main goal of cooperation in highway driving is to ensure that all vehicles in a lane move at the same speed while maintaining the desired formation geometry, specified by a desired inter-vehicle gap policy. The cooperative driving with constant spacing policy, called platooning, mandates vehicles to maintain a constant distance from their immediate predecessor. However, for the cooperative adaptive cruise control (CACC), a constant time headway gap policy is considered, in which the desired following distance should be proportional to the speed of the vehicle; the higher the speed, the larger the distance. The objectives of CACC are to keep the desired small headway time, smooth the engine/brake input, and keep the acceleration within a reasonable and comfortable range. CACC and platooning have the potential to increase the highway capacity when they reach a high market penetration. Studies have shown that a car with a velocity of $80km/h$ following only one predecessor at $25m$ achieves a $30\%$ reduction in aerodynamic drag, and a $40\%$ reduction can be attained by following two predecessors. Improvement in traffic flow for a high market penetration rate is noticeable for platoons as small as three cars. If all passenger cars form vehicle platoons, a $200\%$ growth in the road capacity can be achieved \cite{Impacts_CACC,Traffic_Flow_Characteristics}. It is shown that platooning is more sensitive to communication losses compared to the CACC, mainly due to its very close coupling between vehicles \cite{RZ_impact}. Even though platooning is more sensitive to communication losses, its implementation is stable using Cellular Vehicle-To-Everything (C-V2X) \cite{Razz2204:Finite}.

\begin{figure}
    \centering
    \includegraphics[width=\linewidth]{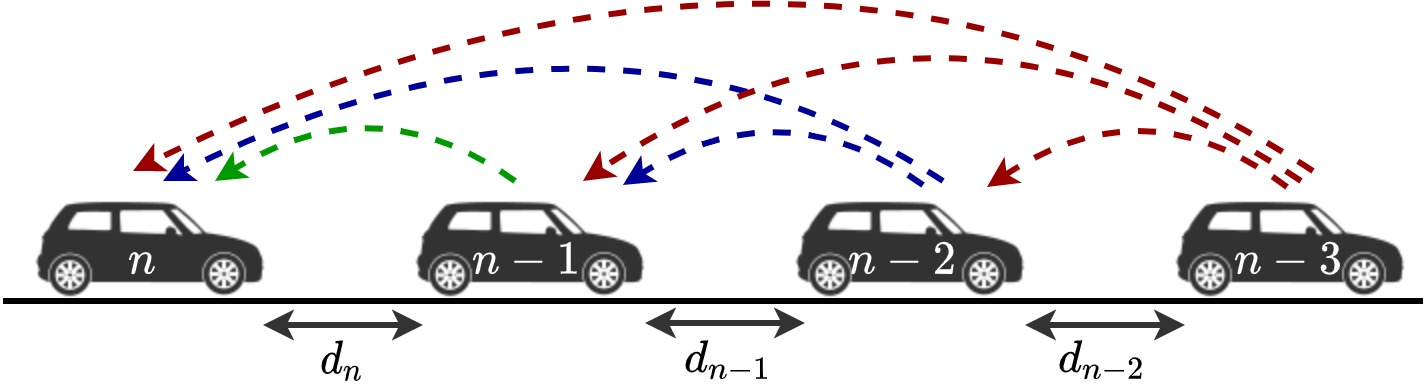}
    \caption{A representation of the communication topology. Dashed lines show the information flow among vehicles. The distance between $i^{th}$ vehicle and its predecessor is shown by $d_i$.}
    \label{fig:diagram}
\end{figure}

Connected and Autonomous Vehicles (CAVs) are required to frequently broadcast their dynamic and kinematic information over the wireless channel. The concept of information sharing among vehicles results in a level of situational awareness for any vehicle and makes it aware of its surrounding environment, crucial for the cooperative safety applications

\restoregeometry
\newgeometry{left=19.1mm,bottom=19.1mm,top=19.1mm,right=19.1mm}
\noindent to function properly\cite{Dataset}. Although the cooperative vehicles’ perception can be expanded to multiple predecessors’ information by using vehicle to vehicle (V2V) communication, these technologies still suffer from scalability issues, especially in congested scenarios. In addition, as a result of packet loss, wireless channels create random link interruption and changes in the network connection. Therefore, it is required for the cooperative vehicles to predict each agent's behavior to compensate for packet loss or latency caused by having non-ideal communication.

Information exchange to create situational awareness is the backbone of distributed systems that rely on communication. The information exchange flow defines how the vehicles in a platoon exchange information with each other. The available information to each controller is often limited to a neighboring region because of the range limitation of sensing and communication systems. As a result, controllers use only local information to achieve a global performance for the platoon. It has been also shown that the time-headway can be minimized using multiple predecessors’ information in the local controller \cite{V2V_time_headway, bian2019reducing, pedro,Influence_IFT,undirected_graph}.

Model-based Communication (MBC) is a recently-explored communication scalability solution, which has shown promising potential to reduce channel congestion \cite{model_based_communication}. The fundamental intention behind the MBC scheme is to utilize a more flexible content structure for broadcasting packets consisting of the joint vehicle/driver behavioral models' parameters in comparison with the Basic Safety Message (BSM) content structure defined by the J2735 standard. For utilizing the MBC scheme, different modeling methods can be considered to represent the vehicle's movement behavior. Non-parametric Bayesian inference techniques, particularly Gaussian Processes (GPs) are amongst the promising methods for analytically tractable modeling of joint vehicle's dynamic/driver's behavior. In addition to exhibiting very good generalization properties, a major advantage of GPs is that they come equipped with a measure of model uncertainty, making them particularly beneficial for safety-critical applications. The driver behavioral models are functions of different factors such as the driver’s driving style, road traffic, weather condition, etc. Therefore, movement models may become very complex. In this research study, a Gaussian Process Regression (GPR) method is used to model cooperative vehicles' velocity trajectories, which allows them to predict the future behavior of their preceding vehicles during communication loss \cite{mosharafian2021gaussian}. 

Cooperative applications need an efficient controller to consider the computational cost and ensure driving comfort and high responsiveness. The advantage of Model Predictive Control (MPC) is that it can realize high control performance since all constraints for these applications can be explicitly dealt with through solving an optimization problem. By utilizing the predicted state of the preceding vehicle, an optimization problem can be solved to obtain the vehicle's control input in MPC-based driving systems. Therefore, the solution provides anticipatory car-following behavior that improves driving efficiency by dynamically adjusting spacing and speed. This paper takes advantage of a discrete hybrid stochastic model predictive control, which incorporates system modes as well as uncertainties captured by GP models. As a safety metric in a vehicle platoon, two operating modes for each vehicle are considered; free following and emergency braking. The proposed control design approach finds the vehicle's optimal velocity trajectory to achieve a safe and efficient platoon of vehicles with a small inter-vehicle gap while reducing the impact of packet loss.

Sudden dynamic changes of vehicles, e.g., hard-braking or shock waves, are the cause of start-and-stop dynamics, which, besides disrupting traffic flow, can lead to accidents \cite{joint_communication/control,stop-and-go}. Traffic shock waves represent a threat in terms of safety and can cause chain collisions when the drivers are distracted or do not respect safety distances. We use a shock wave velocity profile in our simulation studies to examine the effect of Information Flow Topology (IFT) and random packet losses on the safety of a platoon of homogeneous vehicles moving in a rigid formation in shock wave scenarios. We use $0.6s$ headway gap compared to previous works (varying from $0.8s$ to $1.2s$). In our experiments, we study the effect of accessing the information of $r$ leading vehicles. In other words, the ego vehicle can only receive messages from the $r$ preceding vehicles in the platoon ($r$-look-ahead) as shown in Fig. \ref{fig:diagram} for three-look-ahead vehicles. The results show an improvement in response time and reducing the number of emergency braking, demonstrating the potential of cooperation to attenuate disturbances and improve traffic flow. The communication is considered to be non-ideal with varying Packet Error Rates (PERs) and the rate of communication is constant and equal to $10Hz$.


\section{Stochastic Model-based Communication}

\begin{figure*}[h]
\centering
 \includegraphics[width=1\textwidth,trim={5mm 5mm 5mm 5mm},clip]{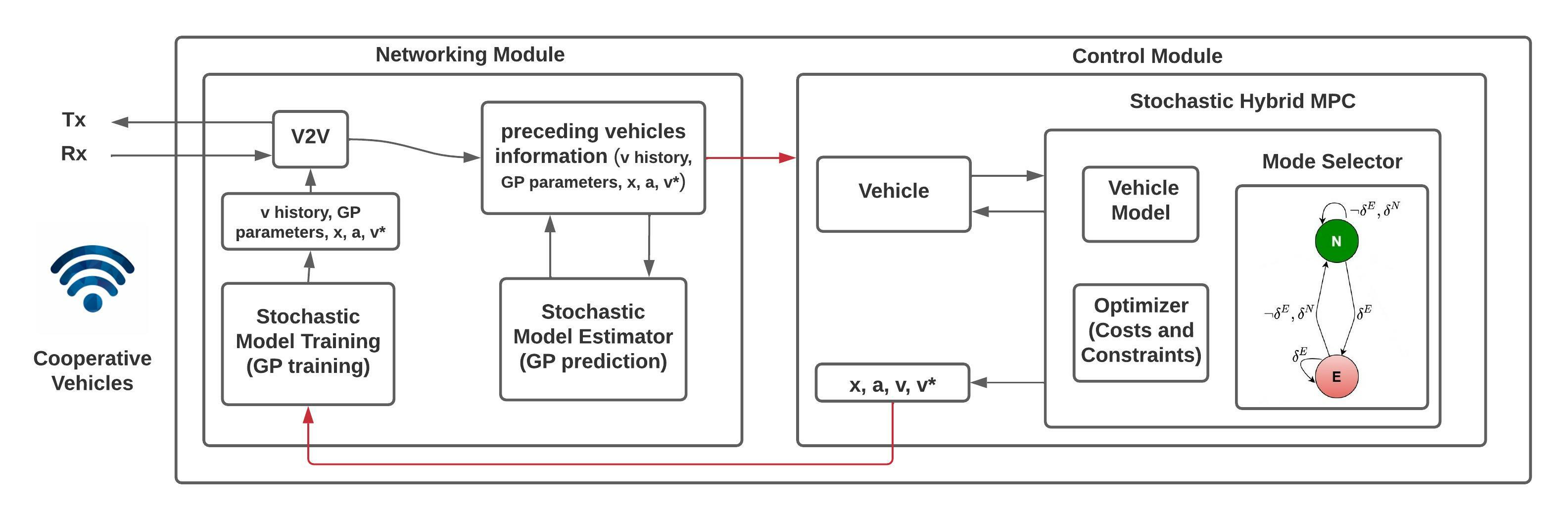}
    \caption{Block diagram of networking and control modules in each member of the platoon. Ego vehicles will receive information from preceding vehicles upon successful communication or will update the information with the stochastic model estimator when the packet is lost. Hybrid stochastic MPC will use the information from the networking module for control purposes. Finally, the control module will pass the current states and velocity predicted values to the networking module for broadcasting.}
    \label{block diagram}
\end{figure*}

\noindent As discussed in Section \ref{sec::intro}, the collective behavior of the platoon is not only dependent on the local controller design but also tightly coupled with the availability and accuracy of the predecessor vehicles' information at each member of the platoon \cite{control_communication_design}. In particular, the stability and scalability of the platoon are dependent on the IFT. To study the effect of IFT on platoon behavior, in this paper, we have considered the $r$-look-ahead directional topology. Although the information shared amongst cooperative vehicles could include precise position, velocity, acceleration, fault warnings, forward hazard warnings, and maximum braking capability, in a vehicle platoon, only the position, velocity, and acceleration information are required. The GP-based MBC trains the GP regression model based on the most recent observed information. This procedure results in generating new situational awareness messages which carry the latest updated abstract model of the vehicle’s state. 

In this paper, we consider the velocity time-series of each cooperative vehicle, $v_{n}(t)$, to be a Gaussian process defined by the mean function $m_{n}(t)$ and the covariance kernel function $\kappa_{n}(t,t^{\prime})$ as
\begin{equation}
v_{n}(\mathbf{t}) \sim \mathcal{G} \mathcal{P}\left(m_{n}(\mathbf{t}), \kappa_{n}\left(\mathbf{t}, \mathbf{t}^{\prime}\right)\right).
\end{equation}
We are interested in incorporating the knowledge that the observed velocity data provides about the underlying function, $v_{n}(t)$, and its future values. 
Assuming that for each cooperative vehicle, the mean of the process is zero, $m_{n}(t)=0$, the covariance kernel is a Radial Basis Function (RBF), and the measurement noises are independent and identically distributed with the Gaussian distribution $\mathcal{N}(0,\,\gamma_{n, noise}^{2})$, the covariance matrix of the observed velocity of the $n^{th}$ cooperative vehicle is
\begin{equation}
\begin{aligned}
\label{kernel matrix}
 K^{n}(\boldsymbol{t},\boldsymbol{t})=K_{r}^{n} + \gamma_{n,noise}^{2}I \\
 \mspace{10mu}[K_r^n]_{ij}=\kappa_{n}(t_i,t_j),
\end{aligned}
\end{equation}
where $I$ denotes the identity matrix of dimension equal to the size of the training (measured) data, and $\kappa_{n}(t_i,t_j)$ can be calculated using the RBF definition as
\begin{equation}
 \kappa_{n}(t,t')=\exp(-\frac{||t-t'||^2}{2\gamma_{n}^{2}}).
\end{equation}
Using the aforementioned assumptions, the joint distribution of the past observed values, $\mathcal{V}_{n}^{obs}$, and the future values $\mathcal{V}_{n}^{\ast}$, can be represented as
\begin{equation}
\left[\begin{array}{l}
\mathbf{\mathcal{V}_{n}^{obs}} \\
\mathbf{\mathcal{V}_{n}}^{*}
\end{array}\right] \sim \mathcal{N}\left(\mathbf{0},\left[\begin{array}{ll}
K^{n}(\boldsymbol{t}, \boldsymbol{t}) & K_{r}^{n}\left(\boldsymbol{t}, \boldsymbol{t^{*}}\right) \\
K_{r}^{n}\left(\boldsymbol{t^{*}}, \boldsymbol{t}\right) & K_{r}^{n}\left(\boldsymbol{t^{*}}, \boldsymbol{t^{*}}\right)
\end{array}\right]\right),
\end{equation}
where $\boldsymbol{t}$ and $\boldsymbol{t^{*}}$ denote the sets of observation and future value time stamps, respectively, and $K^{n}(.,.)$ and $K_{r}^{n}(.,.)$ can be obtained from \eqref{kernel matrix}.
Therefore, the predictive distribution of future velocity values, $\mathcal{V}_{n}^{\ast}$, conditioned on having observed velocity values $\mathcal{V}_{n}^{obs}$ at time stamps $\boldsymbol{t}$ can be derived as
\begin{multline}\label{eq:probabilistic}
    \Big(\mathbf{\mathcal{V}_{n}}^{*} \mid \mathbf{t^{*}}, \mathbf{t}, \mathbf{\mathcal{V}_{n}^{obs}}\Big) \sim \mathcal{N} (\mu_{n}^{\ast},\Sigma_{n}^{\ast}), \\ 
    \mspace{-75mu}\mu_{n}^{\ast}=K_{n}^{r}[\left(t^{*}, t\right)|\alpha_{n}] K_{n}^{-1}[(t, t)|\alpha_{n}] \mathbf{\mathcal{V}_{n}^{obs}}, \\
    \mspace{6mu}\Sigma_{n}^{\ast}=-K_{n}^{r}[\left(t^{*}, t\right)|\alpha_{n}] K_{n}^{-1}[(t, t)|\alpha_{n}]K_{n}^{r}[\left(t, t^{*}\right)|\alpha_{n}] \\
    \mspace{18mu}+K_{n}^{r}[\left(t^{*}, t^{*}\right)|\alpha_{n}].
\end{multline}
Upon each transmission opportunity, each cooperative vehicle uses its $5$ most recent velocity observations, measured at equally-distanced $100 ms$ time intervals, to train a GP model and obtain the set of parameters $\alpha=\{\gamma_{n},\gamma_{n, noise}\}$. For this purpose, the Leave-One-Out Cross Validation has been considered. Assuming that the $i^{th}$ velocity observation, $v_{i}$, is left out, the $\log$ probability of observing $v_{i}$ given the rest of the observations, ($\mathcal{V}_{-i}$), can be represented as
\begin{equation}
\log p\left(v_{i} \mid \mathbf{t}, \mathcal{V}_{-i}, \alpha\right)=-\frac{1}{2} \log \sigma_{i}^{2}-\frac{\left(v_{i}-\mu_{i}\right)^{2}}{2 \sigma_{i}^{2}}-\frac{1}{2} \log 2 \pi,
\end{equation}
where $\mu_{i}$ and $\sigma_{i}$ can be obtained from \eqref{eq:probabilistic}. Defining the cross validation objective function as the sum of the log-likelihoods over all most recent observations, i.e., ${L}(\mathbf{t}, \mathcal{V}, \alpha)=\sum_{i=1}^{5} \log p\left(v_{i} \mid \mathbf{t}, \mathcal{V}_{-i}, \alpha\right)$, the optimal parameters $\alpha^{\ast}_{n}= \{\gamma_{n}^{\ast},\gamma_{n, noise}^{\ast}\}$ can be obtained using the conjugate gradient optimization method as proposed in \cite{Rasmussen:GP}.


After the parameters are learned, the transmitting vehicle shares the model parameters along with its history of the $5$ most recent velocity measurements and current position and acceleration. In addition, $7$ future velocity values (parameter $N$ in Table \ref{table1}) predicted by the vehicle's model predictive controller are included in the transmitting packet. The details of the vehicle's MPC prediction are discussed in the next section. Upon receiving a packet, the receiving vehicle will use the newly received information for local control. Otherwise, in the case of packet loss, the cooperative vehicle will use the GP model to predict the velocity of the transmitting vehicle until receiving a new packet from it using \eqref{eq:probabilistic}.



In addition, the position of the transmitting vehicle is predicted using
\begin{equation}
\begin{aligned}
&\bar{x}_{n}\left(t_{1}\right)=x_{n}\left(t_{0}\right)+\iint_{t_{0}}^{t_{1}} \tilde{v} \operatorname{P}(\tilde{v}) d t d \widetilde{v},
\end{aligned}
\end{equation}
where $P(\tilde{v})$ is the predictive distribution of transmitting vehicle's velocity presented by \eqref{eq:probabilistic}. Therefore, in the case of packet loss, these predicted values will be used for local control.

In our experiments, we considered the packet loss to be an independent and identically distributed random variable and gradually changed from $0\%$ to $60\%$. The communication rate in all experiments is considered to be $10Hz$.

\section{Vehicle Model and Stochastic Model Predictive Control design Approach}

\noindent In this section, the vehicle model and the model predictive control design are explained.

\subsection{Vehicle Model}

In this study, we consider a platoon of $N_v$ vehicles, where $n\in\{0,1,\hdots,N_v\}$ denotes the $n^{th}$ vehicle in the platoon, and $n=0$ represents the platoon leader. As shown in Fig. \ref{fig:diagram}, $d_n$ denotes the gap between $n^{th}$ and $(n-1)^{th}$ vehicles and is defined as
\begin{equation}
    d_n = x_{n-1}-x_{n}-l^v_n,
\end{equation}
where $x_n$ and $l^v_n$ are the longitudinal location of $n^{th}$ vehicle rear bumper and the vehicle length, respectively. We leverage the fixed time headway gap spacing policy, which can improve the string stability and safety\cite{naus2010string}, and define the desired spacing policy as
\begin{equation} \label{d*}
    \begin{aligned}
        d^{*}_{n}(t) = \tau_n\,v_{n}(t)+d^{s}_{n}.
    \end{aligned}
\end{equation}
In \eqref{d*}, $v_{n}$ is the velocity of the $n^{th}$ vehicle, $\tau_n$ is the time gap, and $d^s_n$ represents the standstill distance. The difference between the gap and its desired value is defined as $\Delta d_{n}(t) = d_{n}(t)-d^{*}_{n}(t)$, and the velocity difference between $n^{th}$ vehicle and its predecessor is defined as
$\Delta v_{n}(t) = v_{n-1}(t)-v_{n}(t)$. Hence, $\Delta \dot{d}_n$ turns into $\Delta \dot{d}_n(t)=\Delta v_n(t)-\tau_n\,a_n(t)$ and $\Delta \dot{v}_n=a_{n-1}-a_n$, where $a_n$ denotes the acceleration of the $n^{th}$ vehicle. By taking the driveline dynamics $f_n$ into account, the derivative of the acceleration for vehicle $n$ is $\dot{a_{n}}(t) = -\mathnormal{f}_{n} a_{n}(t) + \mathnormal{f}_{n}u_{n}(t)$, where $u_{n}(t)$ acts as the vehicle input. By considering $S_n=[\Delta d_n\,\,\,\Delta v_n\,\,\, a_n]^T$ as the vector of states for $n^{th}$ vehicle, the state-space representation for each vehicle is
\begin{multline}  \label{css}
        \dot{S}_n(t)=A_n\,S_n(t)+B_n\,u_n(t)+D\,a_{n-1}(t)\\ \\=
        \begin{bmatrix}
            0&1&-\tau_n \\0&0&-1\\0&0&-\mathnormal{f}_{n}
        \end{bmatrix}S_n(t)+
        \begin{bmatrix}
            0\\0\\\mathnormal{f}_{n}
        \end{bmatrix}u_{n}(t)+
        \begin{bmatrix}
            0\\1\\0
        \end{bmatrix}a_{n-1}(t).
\end{multline}
For $n=0$ (leader), $a_{n-1}(t)$ is replaced by zero. The following equation describes the discrete-time state space model when the first-order forward time approximation is employed
\begin{multline}\label{fdss}
        S_n(k+1) =\\ (I+t_s\,A_n)\,S_n(k)+t_s\,B_n\,u_n(k)+t_s\,D\,a_{n-1}(k),
\end{multline}
where $t_s$ is the sampling time.

Some constraints on the system states and input are also considered including bounds on the acceleration and input, road speed limit, and distance between vehicles (note that a negative distance implies collision and therefore should not occur). The following inequalities (hard constraints) should always hold true
\begin{subequations} \label{bounds}
    \begin{gather}
        \label{accbound}
        a_n^{min}\leq a_n(k)\leq a_n^{max},\\ 
        \label{inputbound}
        u_n^{min}\leq u_n(k)\leq u_n^{max},\\
        \label{speedlim}
        v_n(k)\leq v^{max},\\
        d_n(k)>0.
    \end{gather}
\end{subequations}
Besides, for passenger comfort, system input changes are bounded as
\begin{equation}
    \begin{gathered}
        t_s\,u_n^{min} \leq u_n(k+1)-u_n(k) \leq t_s\,u_n^{max}.
    \end{gathered}
    \label{passenger_comfort}
\end{equation}

\subsection{Discrete Hybrid SMPC Design}

Discrete hybrid stochastic automata (DHSA) models a stochastic system with both binary and continuous/discrete-time variables and inputs. DHSA formulation and details are discussed in \cite{bemporad2010model}. Mixed logical dynamical (MLD) form \cite{bemporad1999control} can be used to reformulate a DHSA using linear equations and inequalities. The reformulation enables using mixed-integer programming to find the optimal control input for the system.

In this paper, we consider two operating modes for each vehicle; free following mode and emergency braking mode. In free following mode, each vehicle tries to reach its desired spacing policy while in emergency braking mode, it uses minimum control input and performs hard braking to avoid any possible accident. Emergency braking mode activates when $\Delta d_n(k)$ goes below a fixed level $\underline{d}_n$; in other words, $\Delta d_n(k)+\underline{d}_n \leq 0$. The operating modes add binary variables to the system while employing GPR for predicting velocity turns the system into a stochastic one. To represent the system, DHSA is used, and the system is then reformulated using MLD. The details on how to rewrite each vehicle equations, constraints, and modes in the MLD form can be found in \cite{mosharafian2021gaussian}. After expressing the system in MLD form, the MPC design problem for each vehicle is
\begin{multline} \label{MPC_prob}
        \min_{\textbf{u}_n,\textbf{w}_{n},\textbf{z}_n}\\ \sum_{k=0}^{N-1}\Big[(\mathbf{S}_n(k)-R_n)^T\, Q_n \,(\mathbf{S}_n(k)-R_n)\Big]-q_n\, \ln(\pi(\textbf{w}_{n}))\\
        \text{subject to:}~~~
        \text{MLD system equations,}\\ \ln(\pi(\textbf{w}_{n}))\geq \ln(\tilde{p}_n),~~~~~~~~~~
\end{multline}
where $\textbf{u}_i$ and $\textbf{z}_i$ are the system inputs and the vector of auxiliary variables from $k=0$ to $k=N-1$, respectively, the quadratic term in cost function is the performance index while $q_n\, \ln(\pi(\textbf{w}_{n}))$ is the probability cost, $\textbf{w}_{n}$ stands for all the uncontrollable event variables, which are used to consider the variance of the velocity prediction calculated using GP in the system model, and $\pi(\textbf{w}_{n})$ is the trajectory probability. The last constraint in \eqref{MPC_prob} represents the chance constraint.

\vspace{2mm}

\begin{table}[h]
\caption{Model and optimization parameters used in the simulations.}
\centering
\renewcommand{\arraystretch}{1.5}
\begin{tabular}{||p{1cm}|p{2.45cm}||p{1cm}|p{2.45cm}||} 
 \hline\hline
 parameter & value & parameter & value\\ [0.5ex] 
 \hline\
 $N$ &  $7$ & $t_s$ & $0.1\,s$
 \\\
 $l^v_n$ & $5\,m$ & $d^s_n$ & $2\,m$ \\\
 $\underline{d}_n$ & $0.5\, m$ & $f_n$ & $10\,s^{-1}$\\\
 $a_n^{min}$ & $ -4\,m/s^2$ &  $a_n^{max}$ & $ 3\,m/s^2$\\ 
  $u_n^{min}$ & $ -4\,m/s^2$ &  $u_n^{max}$ & $ 3\,m/s^2$\\ 
 $\tilde{p}_n$ & $0.01^{N}$ & $q_n$ & $10$\\[1ex] 
 \hline\hline
\end{tabular}
\label{table1}
\end{table}

\begin{figure}[t]
  \centering
  \includegraphics[width=1\linewidth]{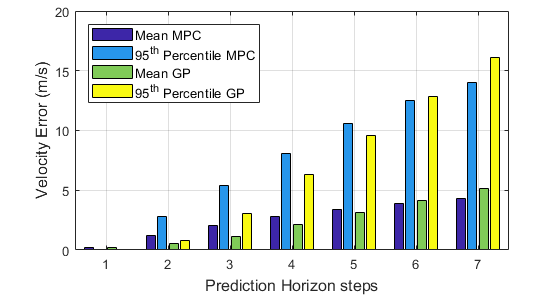}
  \caption{Mean and $95^{th}$ percentile of the absolute velocity error for GP-based prediction and MPC-based prediction for $7$-step ahead prediction horizon using nine-look-ahead topology. Statistics for both methods are almost the same, showing the capability of GP in capturing the velocity profile.}
\label{MPC_GP}
\end{figure}

\begin{figure}[t]
  \centering
    \includegraphics[width=1\linewidth]{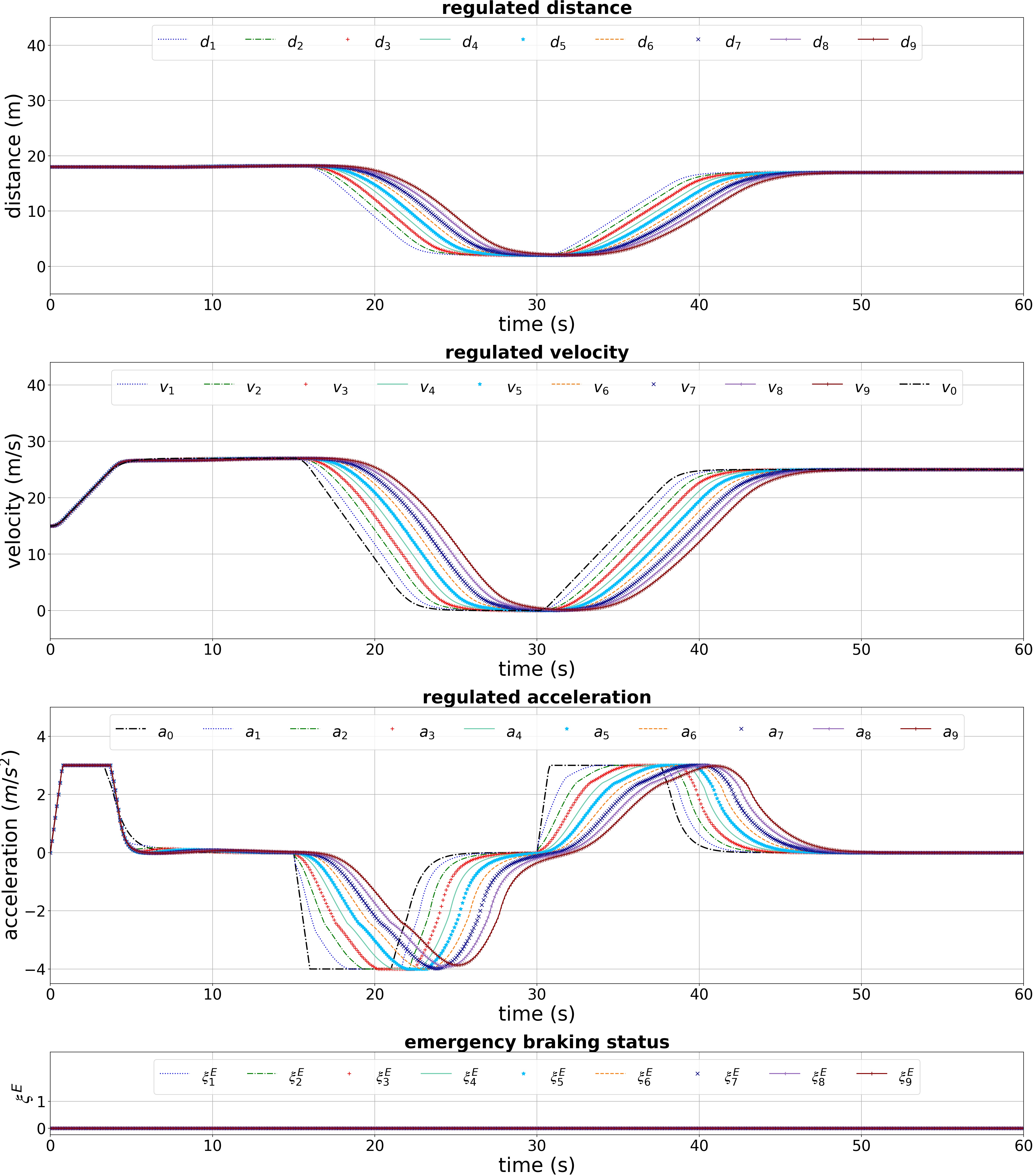}
  \caption{Performance of the CACC with 10 vehicles, $T=0.6\,s$, $t_c=0.1\,s$, and ideal communication with one-look-ahead topology.}
  \label{CACC_ideal}
\end{figure}

\begin{figure*}[pt!]%
\centering
\subcaptionbox{one-look-ahead topology\label{fig:1look}}{%
\includegraphics[width=0.5\textwidth,trim={0mm 0mm 0mm 0mm},clip]{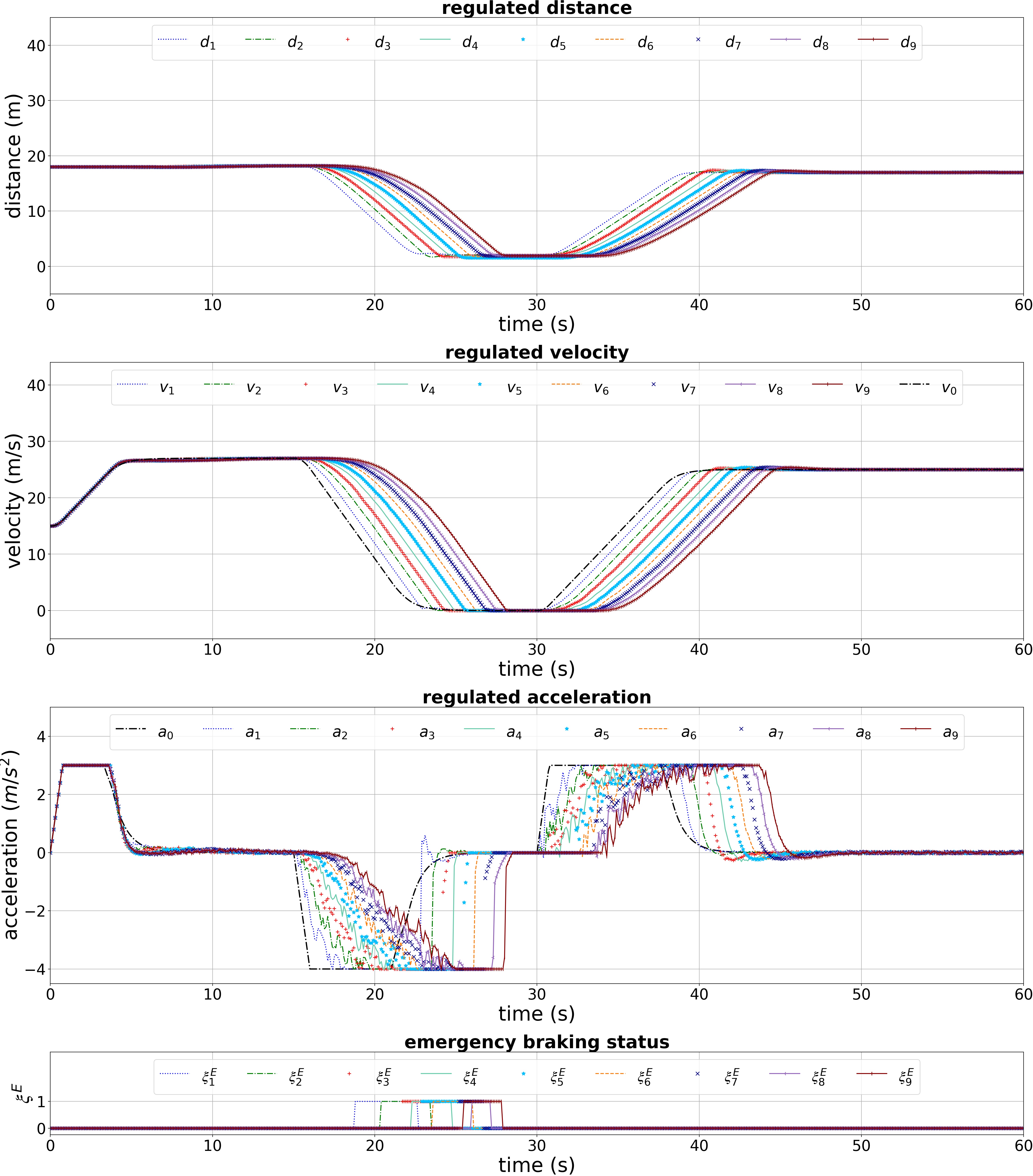}}%
\subcaptionbox{nine-look-ahead topology\label{fig:9look}}{%
\includegraphics[width=.5\textwidth,trim={0mm 0mm 0mm 0mm},clip]{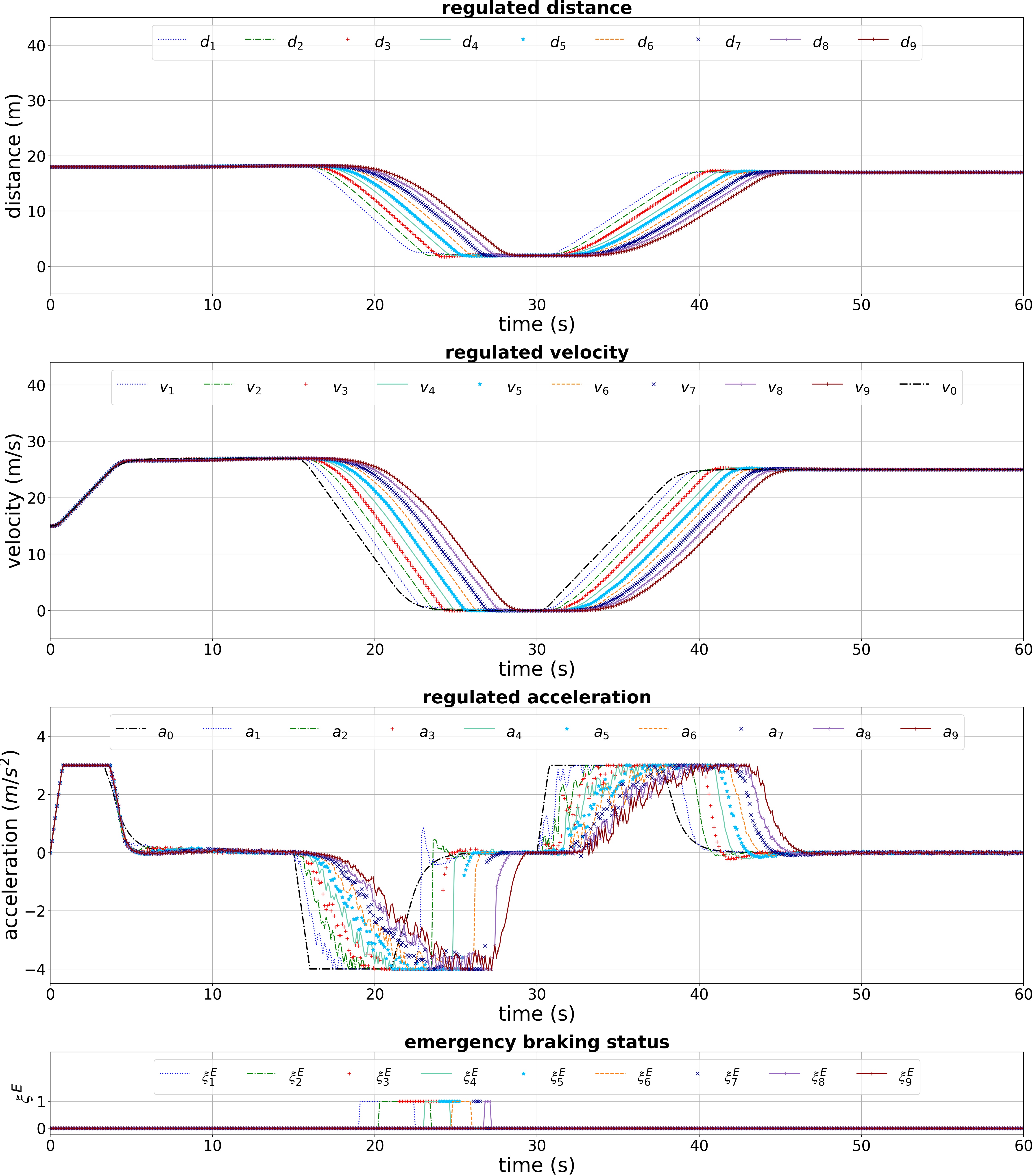}}%
\caption{Performance of the CACC system with 10 vehicles, $T=0.6\,s$, $t_c=0.1\,s$, and $PER = 0.5$.}
\label{fig:CACC_0.5}
\end{figure*}

\begin{remark}
The given MPC problem formulation uses a one-look-ahead topology. The $r$-look-ahead topology can be considered in the problem by changing the cost function in \eqref{MPC_prob} as follows.
\begin{multline}\label{multi_cost}
    \sum_{k=0}^{N-1}\Bigg[(\mathbf{S}_n(k)-R_n)^T\, Q_n \,(\mathbf{S}_n(k)-R_n)\\
    +\hspace{-5pt}\sum_{i=n-r}^{n-1}\hspace{-5pt}\Big[ c^d_i\,\Big(x_i(k)-x_n(k)-\hspace{-15pt}\sum_{j=n-r+1}^{n}\hspace{-10pt}(d^*_j(k)+l^v_j)\Big)^2\\
    +c^v_i\,\Big(v_i(k)-v_n(k)\Big)^2\Big]\Bigg] -q_n\, \ln(\pi(\textbf{w}_{n})),
\end{multline}
where $c^d_i$ and $c^v_i$ are positive coefficients, and $r$ denotes the number of predecessors sharing information with the $n^{th}$ vehicle. In \eqref{multi_cost}, each vehicle tends to achieve the desired distances from its $r$ predecessors while adjusting its velocity based on the predecessors' velocity. It is noted that when $r>n$ (the number of predecessors is less than $r$), the $n^{th}$ vehicle replaces $r$ with $n$ in \eqref{multi_cost}.
\end{remark}

\vspace{2mm}

Each vehicle has access to its preceding vehicle's future velocity trajectory through either the communication (every $t_c$ seconds if packet loss does not occur) or the GP model predictions (until the next successful communication event). During a successful communication event, a vehicle will share its future velocity trajectory (calculated by solving the MPC problem) with its follower vehicles.


\section{Experimental Setup and Simulation Results}

\noindent In our experiments, different $r$-look-ahead one-directional (broadcasting) information flow topologies have been considered for the platoon of cooperative vehicles. For instance, if $r=1$, the receiving vehicle can only receive messages from the nearest predecessor in the platoon (one-look-ahead); however, in other cases, each vehicle can receive messages from several preceding vehicles in the platoon. Moreover, in our experiments, we gradually change the PER values from 0 (ideal communication) to 0.6 (randomly losing 60 percent of packets) to study the effect of communication loss on the CACC performance. The rate of communication is constant and equal to $10Hz$. Simulations are conducted considering a platoon of $10$ vehicles and using multiple scenarios. CVXPY package in Python is used for implementing the optimization problem and Gurobi optimization package is used as the solver for the mixed-integer programming \cite{cvxpy,gurobi}.

The desired velocity trajectory of the leader is set as
\begin{equation}
    \begin{gathered}
    v^*_0(t)=
    \begin{cases}
    27 &\,\,\,\, t<15\,s,\\
    0 &\,\,\,\, 15\,s\leq t < 30\,s,\\
    25 &\,\,\,\, t \geq 30\,s.
    \end{cases}
    \end{gathered}
\end{equation}
The parameters used in the simulations can be found in Table \ref{table1}. Each scenario takes $60s$, in which the objective of the platoon is to maintain the desired gap time of $0.6s$ with the preceding vehicle. Fig. \ref{block diagram} illustrates the overall network/control architecture. The cooperative vehicles update the preceding vehicles' information either based on the newly received information from them or based on the GP predictive model every $100ms$. This information is fed into the hybrid stochastic MPC for updating the control action. In addition, the control module provides the optimal predicted states' values of the ego vehicle. Finally, the current states and the predicted future velocity trajectory will be passed to the networking module for broadcasting.

In order to examine the capability of GPR in modeling the vehicles' velocity trajectories, first, we have designed an experiment in which only the GP models along with the history of the velocity data are shared amongst the vehicles. In this setup, the MPC at each platoon member is fed with the predicted velocities and positions of the preceding vehicles using the shared GP models. Subsequently, the predicted values of cooperative vehicles' velocities for up to $7$ time instants ahead are compared to ground truth to obtain the velocity prediction errors for different time horizons using GPR. In addition, the same procedure has been followed by sharing the MPC states' predictions among the vehicles and the velocity prediction errors using the MPC outputs were derived. Fig. \ref{MPC_GP} illustrates the mean and $95^{th}$ percentile of velocity error for both schemes. It is observed that the statistics for both methods are almost the same, demonstrating the capability of GP for capturing the velocity profile.




\begin{figure}[t]
  \centering
    \includegraphics[width=1.03\linewidth]{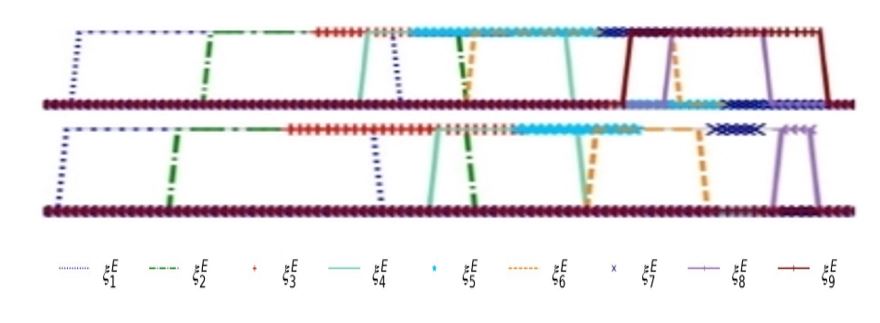}
  \caption{Comparing the emergency braking status for Fig. \ref{fig:1look} with one-look-ahead topology (top) and Fig. \ref{fig:9look} with nine-look-ahead topology (bottom). As observed, using information of more vehicles has led to less emergency braking for platoon members with larger indices (i.e., the ones further down the platoon).}
  \label{zoom}
\end{figure}

\begin{figure}[t]
\centering
\includegraphics[width=1\linewidth]{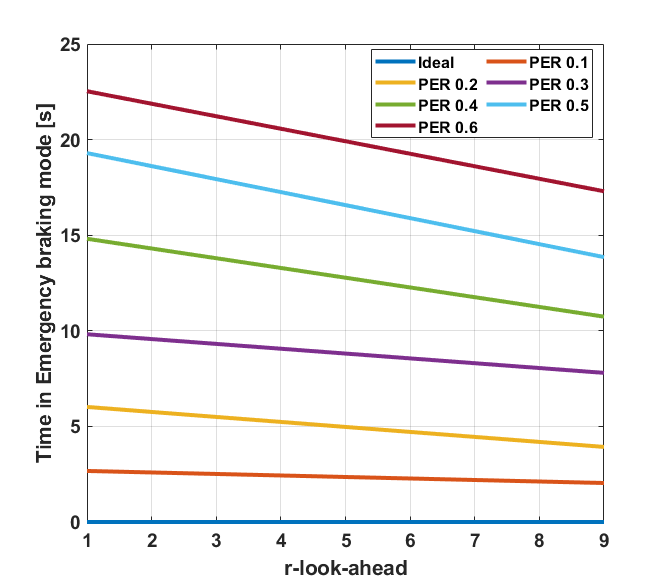}
\caption{The figure shows the sum of emergency braking duration for all platoon members having different look-ahead topologies and PERs. In the case of having semi-ideal communication, using the information of more preceding vehicles does not lead to an evident reduction in emergency braking. However, in normal or adverse situations, the effect of using more vehicles' information in emergency braking duration reduction is more evident.}
\label{EB}
\end{figure}

Two main objectives of CACC are smoothing the engine/brake input, and keeping the acceleration in a reasonable and comfortable range. Therefore, the duration of emergency braking mode and smoothness of the acceleration profile can be considered as two measures for CACC performance evaluation. Fig. \ref{CACC_ideal} shows the regulated distances, velocity and acceleration profiles, and emergency braking status of vehicles in a scenario with ideal communication and one-look-ahead IFT. It is observed that having the information of the immediate predecessor is sufficient to maintain the $0.6s$ headway with smooth acceleration profiles while avoiding the emergency braking given an ideal communication.

In reality, however, the wireless channel is lossy, and the communication can suffer from random packet loss. As the PER increases, it is speculated that by accessing the information of more preceding vehicles, the platoon has a quicker response which leads to the emergency braking reduction. Fig. \ref{fig:CACC_0.5} illustrates the CACC performance when communication PER is set to $0.5$ for one-look-ahead and nine-look-ahead topologies. It is observed that although the vehicles' accelerations similarly fluctuate in both scenarios, using the nine-look-ahead IFT resulted in fewer emergency braking events. Fig. \ref{zoom} shows the emergency braking instances of the platoon members in the aforementioned scenarios in the time interval $[17s,28s]$. It is seen that the emergency braking duration is reduced for almost all platoon members in the nine-look-ahead topology. This effect is more evident for the platoon members with higher indices, i.e., the members at the tail of the platoon. For instance, the last member of the platoon did not have an emergency braking event in the nine-look-ahead scenario while it stayed in the emergency braking mode for a few seconds in the one-look-ahead scenario.    

Fig. \ref{EB} shows the average sum of emergency braking duration of all platoon members having the aforementioned setups and $20$ simulation trials. When the communication is ideal or the PER is relatively small, accessing the information of more predecessors does not extensively reduce the braking emergency duration. As the PER increases, the effect of accessing the information of more preceding vehicles on emergency braking reduction becomes more evident. For instance, when PER is $0.6$, the emergency braking duration is $30\%$ less for nine-look-ahead topology compared to one-look-ahead IFT.

\section{Conclusion}

In this paper, we explored the performance of the vehicle platoon with different information flow topologies, where vehicles receive information from multiple predecessors, in ideal and non-ideal communication setups. In addition, a discrete hybrid stochastic MPC design was used for the CACC application by leveraging model-based communication. It was assumed that vehicles share their future velocity profiles, as well as an updated model for their velocity profile by using GPR at each successful communication event. For safety purposes, vehicles may operate in either free following mode or emergency braking mode. The performance of the proposed controller was evaluated through simulation studies, which validated the efficacy of the proposed method. We have shown that using the information of multiple predecessors can cancel shock waves in a very effective manner. Part of our future work will include the introduction of lane-changing policies for platoons, as lane changing is one of the known causes of shock waves.

\balance
\bibliography{refs.bib}{}
\bibliographystyle{unsrt}

\end{document}